\documentclass[twocolumn]{article}
\usepackage{amsmath}
\usepackage{amsfonts}
\usepackage{achemso}

\input{tcilatex}
\begin{document}

\title{Mean-field electron-vibrational theory of collective effects in
photonic organic materials: bistability}
\author{Boris D. Fainberg \\
Faculty of Sciences, Holon Institute of Technology, 52 Golomb St., Holon
58102, Israel \\
Tel Aviv University, School of Chemistry, Tel Aviv 69978, Israel}
\date{2017/10/26}
\maketitle

\begin{abstract}
Purely organic materials with negative and near-zero dielectric permittivity
can be easily fabricated, and propagation of surface polaritons at the
material/air interface was demonstrated. Here we develop a mean-field theory
of light-induced optical properties of photonic organic materials taking the
collective effects into account. The theory describes both a red shift of
the resonance frequency of isolated molecules, according to the
Clausius-Mossotti Lorentz-Lorentz mechanism, and the wide variations of
their spectra related to the aggregation of molecules into J- or
H-aggregates. We show that the experimental absorption spectra of
H-aggregates may be correctly described only if one takes both mechanisms
into account. The bistable response of organic materials in the condensed
phase has been demonstrated using the electron-vibrational model. We show
that using molecules with long-living triplet state $T_{1}$ near excited
singlet state $S_{1}$, and fast intersystem crossing $S_{1}\rightarrow T_{1}$
enables us to diminish CW light intensity needed for observing bistability
below the damage threshold of thin organic films.
\end{abstract}

\section{Introduction}

Plasmonics and metamaterials provide great scope for concentrating and
manipulating the electromagnetic field on the subwavelength scale to achieve
dramatic enhancement of optical processes and to develop super-resolution
imaging, optical cloaking etc. \cite%
{Stockman08,Nordlander11CR,Maier07,Leonhardt10}. However, metallic
inclusions in metamaterials are sources of strong absorption loss. This
hinders many applications of metamaterials and plasmonics and motivates to
search for efficient solutions to the loss problem \cite{Khurgin15nature}.
Highly doped semiconductors \cite{Hoffman07Nature,Khurgin15nature} and doped
graphene \cite{Abajo11NL,Chen12Nature,Fei12Nature} can in principle solve
the loss problem. However, the plasmonic frequency in these materials is an
order of magnitude lower than that in metals making former most useful at
mid-IR and THz regions. In this relation the question arises whether
metal-free metamaterials and plasmonic systems, which do not suffer from
excessive damping loss, can be realized in the visible range? With no
doubts, inexpensive materials with such advanced properties can impact whole
technological fields of nanoplasmonics and metamaterials.

Recently Noginov et al. demonstrated that purely organic materials
characterized by low losses with negative, near-zero, and smaller than unity
dielectric permittivities can be easily fabricated \cite{Noginov13APL}.
Specifically, the substantially strong negative dielectric permittivity
demonstrated in zinc tetraphenylporphyrin (ZnTPP), suggests that this dye
compound can function as a plasmonic material. The experimental
demonstration of a surface polariton propagating at the ZnTPP/air interface
has been realized \cite{Noginov13APL}.

In addition, Gentile et al. \cite{Gentile14} showed that polymer films doped
with J-aggregated (TDBC) molecules might exhibit a negative real
permittivity in the vicinity of the exciton resonance. Thin films of such
material may support surface exciton-polariton modes, in much the same way
that thin metal films support surface plasmon-polariton modes. Furthermore,
they used the material parameters derived from experiment to demonstrate
that nanostructured excitonic materials may support localized surface
exciton-polariton modes.

Moreover, near-zero dielectric permittivity of the organic host medium
results in dramatic enhancement of intersite dipolar energy-transfer
interaction in the quantum dot wire that influences on electron transport
through nanojunctions \cite{Fainberg15APL}. Such interactions can compensate
Coulomb repulsions in the wire for particular conditions \cite%
{Li_Fai12Nano_Let,White_Fai_Galp12JPCL,Fainberg15APL}. And even the dramatic
laser-induced change of the dielectric permittivity of dyes may be realized\ 
\cite{Noginov15Photonics,Fainberg15APL} that can enable us to control their
"plasmonic" properties.

Both approaches \cite{Noginov13APL} and \cite{Gentile14} were explained in
simple terms of the Lorentz model for linear spectra of dielectric
permittivities of thin film dyes. However, the experiments with strong laser
pulses \cite{Noginov15Photonics} will challenge theory. The point is that
the Lorentz model based on a mean-field theory is described by essentially
nonlinear equations for strong laser excitation. Their general solution is
not a simple problem. In addition, such nonlinear equations can predict
switching waves \cite{Fainberg17APL}, bistability etc.

Here we develop an electron-vibrational theory of light-induced optical
properties of photonic organic materials taking the collective effects into
account. Our consideration is based on the model of the interaction of
strong shaped laser pulse with organic molecules, Refs.\cite%
{Fai98,Fai00JCP,Fai02JCP}, extended to the dipole-dipole intermolecular
interactions in the condensed matter. These latters are taken into account
using a mean-field theory that resulted in two options: one mother - two
daughters. The first option correctly describes the behaviour of the first
moment of molecular spectra in condensed matter, and specifically, the red
shift, according to the Clausius-Mossotti Lorentz-Lorentz (CMLL) mechanism 
\cite{Klein-Furtak86}. The second option is related to the dramatic
modification of molecular spectra in condensed matter due to aggregation of
molecules into J- or H-aggregates. Among other things we demonstrate the
bistable response of organic materials in the condensed phase using the
electron-vibrational model. It is worthy to note that a bistable behavior of
molecular J-aggregates in the context of purely electronic theory was
demonstrated in Refs.\cite{Malyshev96,Malysh2002}. At the same time, as
shown below, the vibrations can give rather important contribution to
broadening aggregate spectra that may be crucial, in spite of strong
narrowing the J-aggregate spectra with respect to the spectra of monomer
molecules. We also consider \ the problem of diminishing light intensity
necessary for bistablity below the damage threshold of thin organic films.

The paper is organized as follows. We start with the derivation of equations
taking dipole-dipole intermolecular interactions in condensed matter into
account. In Sections \ref{sec:CMLL} and \ref{sec:JH-aggregates} we consider
two options of the mean-field theory resulting to the self-energy depending
and not depending, respectively, on effective vibrational coordinate. Then
we describe the absorption of H-aggregates, Section \ref{sec:H}, where we
show that only taking both options of our mean-field theory into account can
explain the experimental results. In Section \ref{sec:bistability} we
consider bistability, and in Section \ref{sec:conclusion}, we briefly
conclude.

\section{Derivation of equations for dipole-dipole intermolecular
interactions in condensed matter}

In this section we shall extend our picture of \textquotedblright
moving\textquotedblright\ potentials of Ref.\cite{Fai98} to a condensed
matter. In this picture we considered a molecule with two electronic states $%
n=1$ (ground) and $2$ (excited) in a solvent described by the Hamiltonian 
\begin{equation}
H_{0}=\sum_{n=1}^{2}|n\rangle \left[ E_{n}+W_{n}(\mathbf{Q})\right] \langle
n|  \label{eq:hamilt}
\end{equation}%
where $E_{2}>E_{1},E_{n}$ is the energy of state $n,W_{n}(\mathbf{Q})$ is
the adiabatic Hamiltonian of reservoir $R$ (the vibrational subsystems of a
molecule and a solvent interacting with the two-level electron system under
consideration in state $n$). The molecule is affected by electromagnetic
field $\mathbf{E}(t)$

\begin{equation}
\mathbf{E}(t)=\frac{1}{2}\mathbf{e}\mathcal{E}(t)\exp (-i\omega t)+\mathrm{%
c.c.}\text{ \ \ }  \label{eq:E_i(t)}
\end{equation}%
the frequency of which is close to that of the transition $1\rightarrow 2$.
Here $\mathcal{E}(t)$ describes the change of the pulse amplitude in time, $%
\mathbf{e}$ is unit polarization vector.

One can describe the influence of the vibrational subsystems of a molecule
and a solvent on the electronic transition within the range of definite
vibronic transition related to a high frequency optically active (OA)
vibration as a modulation of this transition by low frequency (LF) OA
vibrations $\{\omega _{s}\}$ \cite{Fai03AMPS}. In accordance with the
Franck-Condon principle, an optical electronic transition takes place at a
fixed nuclear configuration. Therefore, the quantity $u_{1}(\mathbf{Q}%
)=W_{2}(\mathbf{Q})-W_{1}(\mathbf{Q})-\langle W_{2}(\mathbf{Q})-W_{1}(%
\mathbf{Q})\rangle _{1}$ representing electron-vibration coupling is the
disturbance of nuclear motion under electronic transition where $\langle
\rangle _{n}$ stands for the trace operation over the reservoir variables in
the electronic state $n$. Electronic transition relaxation stimulated by
LFOA vibrations is described by the correlation function $K(t)=\langle
\alpha (0)\alpha (t)\rangle $ of the corresponding vibrational disturbance
with characteristic attenuation time $\tau _{s}$ \cite{Muk95,Fai90OS,Fai93PR}
where $\alpha \equiv -u_{1}/\hbar $. The analytic solution of the problem
under consideration has been obtained due to the presence of a small
parameter. For broad vibronic spectra satisfying the \textquotedblright slow
modulation\textquotedblright\ limit, we have $\sigma _{2s}\tau _{s}^{2}\gg 1$
where $\sigma _{2s}=K(0)$ is the LFOA vibration contribution to a second
central moment of an absorption spectrum, the half bandwidth of which is
related to $\sigma _{2s}$ as $\Delta \omega _{abs}=2\sqrt{2\sigma _{2s}\ln 2}
$. According to Refs. \cite{Fai90OS,Fai93PR}, the following times are
characteristic for the time evolution of the system under consideration: $%
\sigma _{2s}^{-1/2}<T^{\prime }<<\tau _{s}$, where $\sigma _{2s}^{-1/2}$ and 
$T^{\prime }=(\tau _{s}/\sigma _{2s})^{1/3}$ are the times of reversible and
irreversible dephasing of the electronic transition, respectively. The
characteristic frequency range of changing the optical transition
probability can be evaluated as the inverse $T^{\prime }$, i.e. $(T^{\prime
})^{-1}.$ Thus, one can consider $T^{\prime }$ as a time of the optical
electronic transition. Therefore, the inequality $\tau _{s}\gg T^{\prime }$
implies that the optical transition is instantaneous where relation $%
T^{\prime }/\tau _{s}<<1$ plays the role of a small parameter.

This made it possible to describe vibrationally non-equilibrium populations
in electronic states $1$ and $2$ by balance equations for the intense pulse
excitation (pulse duration $t_{p}>T^{\prime }$) and solve the problem \cite%
{Fai90CP,Fai98}.

Let us include now the dipole-dipole intermolecular interactions in the
condensed matter that are described by Hamiltonian \cite{Dav71,Muk95} $%
H_{int}=\hbar \sum\limits_{m\neq n}J_{mn}b_{m}^{\dag }b_{n}$. Then Eq.(6) of
Ref. \cite{Fai98} describing vibrationally non-equilibrium populations in
electronic states $j=1,2$ for the exponential correlation function $%
K(t)/K(0)\equiv S(t)=\exp (-|t|/\tau _{s})$ can be written as%
\begin{eqnarray}
\frac{\partial }{\partial t}\rho _{jj}\left( \alpha ,t\right) &=&-i\hbar
^{-1}[H_{0}\left( \alpha ,t\right) +H_{int}-\mathbf{D}\cdot \mathbf{E}\left(
t\right) ,\rho \left( \alpha ,t\right) ]_{jj}  \notag \\
&&+L_{jj}\rho _{jj}\left( \alpha ,t\right)  \label{eq:rhojj}
\end{eqnarray}%
where $j=1,2;$ and the operator $L_{jj}$ is determined by the equation: 
\begin{equation}
L_{jj}=\tau _{s}^{-1}\left[ 1+\left( \alpha -\delta _{j2}\omega _{st}\right) 
\frac{\partial }{\partial \left( \alpha -\delta _{j2}\omega _{st}\right) }%
+\sigma _{2s}\frac{\partial ^{2}}{\partial \left( \alpha -\delta _{j2}\omega
_{st}\right) ^{2}}\right] ,  \label{eq:Ljj}
\end{equation}%
describes the diffusion with respect to the coordinate $\alpha $ in the
corresponding effective parabolic potential $U_{j}\left( \alpha \right) $, $%
\delta _{ij}$ is the Kronecker delta, $\omega _{st}=\beta \hbar \sigma _{2s}$
is the Stokes shift of the equilibrium absorption and luminescence spectra, $%
\beta =1/k_{B}T$. In the absence of the dipole-dipole intermolecular
interactions in the condensed matter, $H_{int}$, Eq.(\ref{eq:rhojj}) is
reduced to Eq.(11) of Ref.\cite{Fai98}. The partial density matrix of the
system $\rho _{jj}\left( \alpha ,t\right) $ describes the system
distribution in states $1$ and $2$ with a given value of $\alpha $ at time $%
t $. The complete density matrix averaged over the stochastic process which
modulates the system energy levels, is obtained by integration of $\rho
_{ij}\left( \alpha ,t\right) $ over $\alpha $, $\langle \rho \rangle
_{ij}\left( t\right) =\int \rho _{ij}\left( \alpha ,t\right) d\alpha $,
where quantities $\langle \rho \rangle _{jj}\left( t\right) $ are the
normalized populations of the corresponding electronic states: $\langle \rho
\rangle _{jj}\left( t\right) \equiv n_{j}$, $n_{1}+n_{2}=1$. Knowing $\rho
_{jj}\left( \alpha ,t\right) $, one can calculate the positive frequency
component of the polarization $\mathbf{P}^{(+)}(t)=N\mathbf{D}_{12}\langle
\rho \rangle _{21}\left( t\right) $, and the susceptibility $\chi (\Omega
,t) $ \cite{Fai98} that enables us to obtain the dielectric function $%
\varepsilon $ due to relation $\varepsilon (\Omega ,t)=1+4\pi \chi (\Omega
,t)$. Here $N$ is the density of molecules. It is worthy to note that
magnitude $\varepsilon (\Omega ,t)$ does make sense, since it changes in
time slowly with respect to dephasing. In other words, $\varepsilon (\Omega
,t)$ changes in time slowly with respect to reciprocal characteristic
frequency domain of changing $\varepsilon (\Omega )$.

Consider the contribution of Hamiltonian $\hat{H}_{int}$ to the change of $%
\rho _{ij}\left( \alpha ,t\right) $ in time. In other words, we shall
generalize Eq.(11) of Ref.\cite{Fai98} to the dipole-dipole intermolecular
interactions in the condensed matter. Using the Heisenberg equations of
motion, one obtains that $\hat{H}_{int}$ gives the following contribution to
the change of the expectation value of excitonic operator $b_{k}$ in time%
\begin{eqnarray}
\frac{d}{dt}\langle b_{k}\rangle &\sim &\frac{i}{\hbar }\langle \lbrack \hat{%
H}_{int},b_{k}]\rangle \equiv \frac{i}{\hbar }Tr([\hat{H}_{int},b_{k}]\rho )
\notag \\
&=&-i\sum\limits_{n\neq k}J_{kn}\langle (\hat{n}_{k1}-\hat{n}%
_{k2})b_{n}\rangle  \label{eq:Heis1}
\end{eqnarray}%
where $\hat{n}_{k1}=b_{k}b_{k}^{\dag }$, and $\hat{n}_{k2}=b_{k}^{\dag
}b_{k} $ is the exciton population operator. Considering an assembly of
identical molecules, one can write $\langle b_{k}\rangle =\rho _{21}\left(
\alpha ,t\right) $ \cite{Fainberg10APC} if averaging in Eq.(\ref{eq:Heis1})
is carried out using density matrix $\rho \left( \alpha ,t\right) $.
Consider the expectation value $\langle (\hat{n}_{k1}-\hat{n}%
_{k21})b_{n}\rangle =Tr[(\hat{n}_{k1}-\hat{n}_{k2})b_{n}\rho \left( \alpha
_{k},\alpha _{n},t\right) ] $ for $n\neq k$ where $\alpha _{m}$ is the
effective vibrational coordinate of a molecule $m$ ($m=k,n$). Due to fast
dephasing (see above), it makes sense to neglect all correlations among
different molecules \cite{Muk95}, and set $\langle (\hat{n}_{k1}-\hat{n}%
_{k2})b_{n}\rangle =\langle \hat{n}_{k1}-\hat{n}_{k2}\rangle \langle
b_{n}\rangle $ and correspondingly $\rho \left( \alpha _{k},\alpha
_{n},t\right) \simeq \rho \left( \alpha _{k},t\right) \rho \left( \alpha
_{n},t\right) $, i.e. density matrix $\rho \left( \alpha _{k},\alpha
_{n},t\right) $ is factorized. Here from dimension consideration one
expectation value should be calculated using density matrix $\rho \left(
\alpha ,t\right) $, and another one - using $\langle \rho \rangle \left(
t\right) =\int \rho \left( \alpha ,t\right) d\alpha $. Since we sum with
respect to $n$, it would appear reasonable to integrate with respect to $%
\alpha _{n}$. However, this issue is not so simple. The point is that in
addition to intramolecular vibrations, there is a contribution of
low-frequency intermolecular and solvent coordinates into effective
coordinate $\alpha $. Because of this, partitioning the vibrations into $%
\alpha _{k}$ and $\alpha _{n}$ groups is ambiguous, and the mean-field
approximation gives two options

\begin{equation}
p\langle b\rangle \langle \hat{n}_{1}-\hat{n}_{2}\rangle =\binom{p\rho
_{21}\left( \alpha ,t\right) \Delta n}{p\langle \rho _{21}\rangle (t)\Delta
^{\prime }\left( \alpha ,t\right) }  \label{eq:SE-p1}
\end{equation}%
where $\Delta ^{\prime }\left( \alpha ,t\right) =\rho _{11}\left( \alpha
,t\right) -\rho _{22}\left( \alpha ,t\right) $, $p\equiv -\sum_{n\neq
k}J_{kn}$, $\Delta n\equiv n_{1}-n_{2}$. Below we shall discuss which option
better corresponds to a specific experimental situation. Consideration based
on non-equilibrium Green functions (GF) shows that the terms $p\rho
_{21}\left( \alpha ,t\right) $ and $p\langle \rho _{21}\rangle (t)$ on the
right-hand-side of Eq.(\ref{eq:SE-p1}) represent the self-energy, $\Sigma
_{21}(t)$, and the terms $\Delta n$ and $\Delta ^{\prime }\left( \alpha
,t\right) $ - the difference of the "lesser" GFs for equal time arguments, $%
G_{11}^{<}(t,t)-G_{22}^{<}(t,t)$, that are the density matrix, i.e. $%
p\langle b\rangle \langle \hat{n}_{1}-\hat{n}_{2}\rangle =\Sigma
_{21}(t)[G_{11}^{<}(t,t)-G_{22}^{<}(t,t)]$, respectively. In other words,
for the first line on the right-hand-side of Eq.(\ref{eq:SE-p1}), the
self-energy depends on $\alpha $ and the "lesser" GFs $%
G_{11}^{<}(t,t)-G_{22}^{<}(t,t)$ do not. In contrast, for the second line on
the right-hand-side of Eq.(\ref{eq:SE-p1}), the self-energy does not depend
on $\alpha $ and the "lesser" GFs $G_{11}^{<}(\alpha ;t,t)-G_{22}^{<}(\alpha
;t,t)$ do depend. This yields $\partial \rho _{21}\left( \alpha ,t\right)
/\partial t\sim i\Sigma _{21}(t)[G_{11}^{<}(t,t)-G_{22}^{<}(t,t)]$. Adding
term "$i\Sigma _{21}(t)[G_{11}^{<}(t,t)-G_{22}^{<}(t,t)]$" to the right-hand
side of Eq.(9) of Ref.\cite{Fai98}

\begin{eqnarray}
&&\frac{\partial }{\partial t}\tilde{\rho}_{21}\left( \alpha ,t\right)
+i\left( \omega _{21}-\omega -\alpha \right) \tilde{\rho}_{21}\left( \alpha
,t\right)  \notag \\
&\approx &\frac{i}{2\hbar }\mathbf{D}_{21}\cdot \mathbf{E}\left( t\right)
\Delta ^{\prime }\left( \alpha ,t\right) +i\tilde{\Sigma}%
_{21}(t)[G_{11}^{<}(t,t)  \notag \\
&&-G_{22}^{<}(t,t)]  \label{eq:rho21}
\end{eqnarray}%
where $\tilde{\rho}_{21}=$ $\rho _{21}\exp \left( i\omega t\right) $, $%
\tilde{\Sigma}_{21}=$ $\Sigma _{21}\exp \left( i\omega t\right) ,$ and using
the procedure described there, we get the extensions of Eq.(11) of Ref.\cite%
{Fai98} to the dipole-dipole intermolecular interactions in the condensed
matter.

\section{Self-energy depending on effective vibrational coordinate $\protect%
\alpha $}

\label{sec:CMLL}

Consider first the case of self-energy depending on effective vibrational
coordinate $\alpha $ (the first line on the right-hand-side of Eq.(\ref%
{eq:SE-p1})) when the main contribution to $\alpha $ is due to low-frequency
intermolecular vibrations and solvent coordinates. Then we arrive to
equation 
\begin{eqnarray}
\frac{\partial \rho _{jj}\left( \alpha ,t\right) }{\partial t} &=&L_{jj}\rho
_{jj}\left( \alpha ,t\right) +\frac{\left( -1\right) ^{j}\pi }{2}\Delta
^{\prime }\left( \alpha ,t\right) |\Omega _{R}(t)|^{2}\times  \notag \\
&&\times \delta \lbrack \omega _{21}-p\Delta n-\omega -\alpha ]
\label{eq:rhojjfin_b}
\end{eqnarray}%
where $\omega _{21}$ is the frequency of Franck-Condon transition $%
1\rightarrow 2$, $\Omega _{R}(t)=(\mathbf{D}_{12}\cdot \mathbf{e})\mathcal{E}%
(t)/\hbar $ is the Rabi frequency, $\mathbf{D}_{12}$ is the electronic
matrix element of the dipole moment operator.

As one can see from Eq.(\ref{eq:rhojjfin_b}), self-energy $\Sigma
_{21}(t)=p\rho _{21}\left( \alpha ,t\right) $ results in the frequency shift
of spectra "$-p\Delta n$" without changing the line shapes \cite%
{Fainberg15APL}. One can show that this approach correctly describes the
change of the first moment of optical spectra in the condensed matter.
Calculations of $p$ for isotropic medium give $p=\dfrac{4\pi }{3\hbar }%
|D_{12}|^{2}N>0$ \cite{Bowden92,Muk95,Fainberg15APL} that corresponds to a
red shift, according to the Clausius-Mossotti Lorentz-Lorentz (CMLL)
mechanism \cite{Klein-Furtak86}.

Integration of Eq.(\ref{eq:rhojjfin_b}) is achieved by the Green's function 
\cite{Fai90CP} 
\begin{align}
G_{jj}\left( \alpha ,t;\alpha ^{\prime },t^{\prime }\right) & =\frac{1}{%
\sqrt{2\pi \sigma \left( t-t^{\prime }\right) }}\exp \{-[\left( \alpha
-\delta _{j2}\omega _{st}\right)  \notag \\
& -\left( \alpha ^{\prime }-\delta _{j2}\omega _{st}\right) S\left(
t-t^{\prime }\right) ]^{2}/\left( 2\sigma \left( t-t^{\prime }\right)
\right) \}  \label{eq:GF}
\end{align}%
where $\sigma \left( t-t^{\prime }\right) =\sigma _{2s}\left[ 1-S^{2}\left(
t-t^{\prime }\right) \right] $, for the initial condition,

\begin{equation}
\rho _{jj}^{\left( 0\right) }\left( \alpha \right) \equiv \rho _{jj}\left(
\alpha ,t=0\right) =\delta _{j1}\left( 2\pi \sigma _{2s}\right) ^{-1/2}\exp
(-\frac{\alpha ^{2}}{2\sigma _{2s}})  \label{eq:IC}
\end{equation}%
We obtain 
\begin{eqnarray}
\rho _{jj}\left( \alpha ,t\right) &=&\rho _{jj}^{\left( 0\right) }\left(
\alpha \right) +\left( -1\right) ^{j}\frac{\pi }{2}  \notag \\
&&\times \int_{0}^{t}dt^{\prime }|\Omega _{R}(t^{\prime })|^{2}\Delta
^{\prime }\left( \omega _{21}-p\Delta n-\omega ,t^{\prime }\right)  \notag \\
&&\times G_{jj}\left( \alpha ,t;\omega _{21}-p\Delta n-\omega ,t^{\prime
}\right)  \label{eq:rhojjint2}
\end{eqnarray}%
where $\Delta ^{\prime }\left( \omega _{21}-p\Delta n-\omega ,t^{\prime
}\right) $ satisfies nonlinear integral equations that can be easily
obtained from Eq.(\ref{eq:rhojjint2}). Integrating both sides of Eq.(\ref%
{eq:rhojjint2}) with respect to $\alpha $ and bearing in mind that

$\int_{-\infty }^{\infty }G_{jj}\left( \alpha ,t;\omega _{21}-\omega \left(
t^{\prime }\right) ,t^{\prime }\right) d\alpha =1,$ we get%
\begin{equation}
\frac{dn_{j}}{dt}=\left( -1\right) ^{j}\frac{\pi }{2}|\Omega
_{R}(t)|^{2}\Delta ^{\prime }\left( \omega _{21}-p\Delta n-\omega ,t\right)
\label{eq:n_j2}
\end{equation}

\subsection{Fast vibrational relaxation}

Let us consider the particular case of fast vibrational relaxation when one
can put the normalized correlation function $S\left( t-t^{\prime }\right)
\equiv K\left( t-t^{\prime }\right) /K\left( 0\right) $ equal to zero.
Physically it means that the equilibrium distributions into the electronic
states have had time to be set during changing the pulse parameters. Bearing
in mind that for fast vibronic relaxation

\begin{eqnarray}
\Delta ^{\prime }\left( \alpha ,t\right) &=&\frac{n_{1}\left( t\right) }{%
\left( 2\pi \sigma _{2s}\right) ^{1/2}}\exp (-\frac{\alpha ^{2}}{2\sigma
_{2s}})-  \notag \\
&&-\frac{n_{2}\left( t\right) }{\left( 2\pi \sigma _{2s}\right) ^{1/2}}\exp
[-\frac{(\alpha -\omega _{st})^{2}}{2\sigma _{2s}}],  \label{eq:Delta'eq}
\end{eqnarray}%
substituting the last equation into Eq.(\ref{eq:n_j2}) and using Eq.(\ref%
{eq:rho21integrated1}), one gets 
\begin{eqnarray}
\frac{dn_{j}}{dt} &=&\left( -1\right) ^{j}\sigma _{a}(\omega _{21})\tilde{J}%
(t)\func{Re}[n_{1}\bar{W}_{a}(\omega +p\Delta n)-  \notag \\
&&-n_{2}\bar{W}_{f}(\omega +p\Delta n)]-\left( -1\right) ^{j}\frac{n_{2}}{%
T_{1}}  \label{eq:n_j5}
\end{eqnarray}%
where $n_{1}+n_{2}=1$, $\sigma _{a}$ is the cross section at the maximum of
the absorption band, $\tilde{J}(t)$ is the power density of exciting
radiation, $\bar{W}_{a(f)}(\omega )=W_{a(f)}(\omega )/F_{a,\max }$, and we
added term "$\left( -1\right) ^{j}n_{2}/T_{1}$" taking the lifetime $T_{1}$
of the excited state into account. Here "$-iW_{a(f)}(\omega )$" is the
line-shape function of a monomer molecule for the absorption (fluorescence)
for fast vibronic relaxation, and

\begin{eqnarray}
&&\int_{-\infty }^{\infty }d\alpha \Delta ^{\prime }\left( \alpha ,t\right)
\zeta (\omega -\omega _{21}+\alpha )/\pi  \notag \\
&=&-i[n_{1}\left( t\right) W_{a}(\omega )-n_{2}\left( t\right) W_{f}(\omega
)]  \label{eq:lineshape1}
\end{eqnarray}%
where $\zeta (\omega -\omega _{21}+\alpha )=\frac{P}{\omega -\omega
_{21}+\alpha }-i\pi \delta (\omega -\omega _{21}+\alpha )$, $P$ is the
symbol of the principal value.

The imaginary part of "$-iW_{a(f)}(\omega )$" with sign minus, $-\func{Im}%
[-iW_{a(f)}(\omega )]=\func{Re}W_{a(f)}(\omega )\equiv F_{a(f)}(\omega )$,
describes the absorption (fluorescence) lineshapes of a monomer molecule,
and the real part, $\func{Re}[-iW_{a(f)}(\omega )]=\func{Im}W_{a(f)}(\omega
) $, describes the corresponding refraction spectra.

For the \textquotedblright slow modulation\textquotedblright\ limit
considered in this section, 
\begin{equation}
W_{a(f)}(\omega )=\sqrt{\frac{1}{2\pi \sigma _{2s}}}w(\frac{\omega -\omega
_{21}+\delta _{a(f),f}\omega _{st}}{\sqrt{2\sigma _{2s}}})  \label{eq:W}
\end{equation}%
where $w(z)=\exp (-z^{2})[1+i\func{erf}i(z)]$ is the probability integral of
a complex argument \cite{Abr64}, and 
\begin{equation}
F_{a(f)}(\omega )=\sqrt{\frac{1}{2\pi \sigma _{2s}}}\exp [-\frac{\left(
\omega _{21}-\omega -\delta _{a(f),f}\omega _{st}\right) ^{2}}{2\sigma _{2s}}%
]  \label{eq:abs}
\end{equation}

\section{Population difference ("lesser" GFs) depending on effective
vibrational coordinate $\protect\alpha $}

\label{sec:JH-aggregates}

Consider now the case when the population difference depends on effective
vibrational coordinate $\alpha $ (the second line on the right-hand-side of
Eq.(\ref{eq:SE-p1}); the main contribution to $\alpha $ is due to
intramolecular vibrations). Then we arrive to equation%
\begin{eqnarray}
\frac{\partial \rho _{jj}\left( \alpha ,t\right) }{\partial t} &=&L_{jj}\rho
_{jj}\left( \alpha ,t\right) +\frac{\left( -1\right) ^{j}\pi }{2}\Delta
^{\prime }\left( \alpha ,t\right)  \notag \\
&&\times \delta \left( \omega _{21}-\omega -\alpha \right) |\Omega
_{eff}(t)|^{2}  \label{eq:rhojjfin_b2}
\end{eqnarray}%
where $\Omega _{eff}(t)=\Omega _{R}(t)+2p\langle \rho _{21}\rangle
(t)=\Omega _{R}(t)+2\Sigma _{21}(t)$ is the effective Rabi frequency that
can be written as 
\begin{equation}
\Omega _{eff}(t)=\frac{\Omega _{R}(t)}{1+p\int d\alpha \Delta ^{\prime
}\left( \alpha ,t\right) \zeta (\omega -\omega _{21}+\alpha )},
\label{eq:rho21integrated1}
\end{equation}%
Here $\int d\alpha \Delta ^{\prime }\left( \alpha ,t\right) \zeta (\omega
-\omega _{21}+\alpha )/\pi $ is the line-shape function that is reduced to
that of a monomer molecule in the absense of the dipole-dipole interactions
for the equilibrium value of $\Delta ^{\prime }\left( \alpha ,t\right) $.

One can see that in contrast to the self-energy depending on effective
vibrational coordinate $\alpha $, Section \ref{sec:CMLL}, here the
self-energy $\Sigma _{21}(t)=p\langle \rho _{21}\rangle (t)$ (the second
line on the right-hand-side of Eq.(\ref{eq:SE-p1})) results in the change of
both the frequency shift of spectra and their lineshapes. In that case
considering the dense collection of molecules under the action of one more
(weak) field

\begin{equation*}
\mathbf{\tilde{E}}(t)=\frac{1}{2}\mathbf{e}\mathcal{\tilde{E}}(t)\exp
(-i\Omega t)+\mathrm{c.c.,}\text{ \ \ }
\end{equation*}%
one gets for the positive frequency component of the polarization $\mathbf{P}%
^{+}=N\mathbf{D}_{12}\langle \rho _{21}\rangle (t)$%
\begin{equation}
\mathbf{P}^{+}(\Omega ,t)=\frac{-N\mathbf{D}_{12}(\mathbf{D}_{21}\cdot 
\mathbf{e})\mathcal{\tilde{E}}(t)/(6\hbar )}{[\int d\alpha \Delta ^{\prime
}\left( \alpha ,t\right) \zeta (\Omega -\omega _{21}+\alpha )]^{-1}+p},
\label{eq:P^+(t)a}
\end{equation}%
for the susceptibility 
\begin{equation}
\chi (\Omega ,t)=-\dfrac{N|D_{12}|^{2}}{3\hbar }\frac{\int d\alpha \Delta
^{\prime }\left( \alpha ,t\right) \zeta (\Omega -\omega _{21}+\alpha )}{%
1+p\int d\alpha \Delta ^{\prime }\left( \alpha ,t\right) \zeta (\Omega
-\omega _{21}+\alpha )}  \label{eq:suc}
\end{equation}%
and the dielectric function%
\begin{eqnarray}
\varepsilon (\Omega ,t) &=&1-4\pi \dfrac{N|D_{12}|^{2}}{3\hbar }  \notag \\
&&\times \frac{\int d\alpha \Delta ^{\prime }\left( \alpha ,t\right) \zeta
(\Omega -\omega _{21}+\alpha )}{1+p\int d\alpha \Delta ^{\prime }\left(
\alpha ,t\right) \zeta (\Omega -\omega _{21}+\alpha )}  \label{eq:dielectric}
\end{eqnarray}

It is worthy to note that the \textquotedblright slow
modulation\textquotedblright\ limit underlying Eq.(\ref{eq:rhojjfin_b2}) can
break down in the case of the formation of J-aggregates possessing rather
narrow spectra. In such a case one should use more general theory that is
not based on the approximation of broad vibronic spectra. Below we shall
consider the case of fast vibrational relaxation that is not limited by the
\textquotedblright slow modulation\textquotedblright .

\subsection{Line-shape in the fast vibrational relaxation limit}

\label{sec:lineshape}

We shall see below that the approximation based on the self-energy
integrated on the effective vibrational coordinate (the second line on the
right-hand-side of Eq.(\ref{eq:SE-p1})) correctly describe the exciton
spectra. In this relation, the fast vibrational relaxation limit for the
case under consideration should be used with caution. The point is that the
equilibrium state under study is rather the equilibrium state of the
collective system (molecules coupled by the dipole-dipole interaction) than
that of separate molecules. However, the exciton wave function in the ground
state is the product of the wave functions of monomers \cite{Dav71} (no
intermolecular interactions). Because of this, Eq.(\ref{eq:rhojjfin_b2}) for
the absorption of weak radiation and $j=1$ can be written as

\begin{eqnarray}
\frac{\partial \rho _{11}\left( \alpha ,t\right) }{\partial t} &=&L_{11}\rho
_{11}\left( \alpha ,t\right) -\frac{\pi }{2}\Delta ^{\prime (0)}\left(
\alpha ,t\right)  \notag \\
&&\times \delta \left( \omega _{21}-\omega -\alpha \right) |\Omega
_{eff}^{(0)}(t)|^{2}  \label{eq:rhojjfin_b3}
\end{eqnarray}%
where%
\begin{equation}
\Omega _{eff}^{(0)}(t)=\frac{\Omega _{R}(t)}{1+p\int d\alpha \Delta ^{\prime
(0)}\left( \alpha \right) \zeta (\omega -\omega _{21}+\alpha )},
\label{eq:rho21integrated_eq}
\end{equation}%
$\Delta ^{\prime (0)}\left( \alpha \right) =\rho _{11}^{\left( 0\right)
}\left( \alpha \right) =\left( 2\pi \sigma _{2s}\right) ^{-1/2}\exp [-\alpha
^{2}/(2\sigma _{2s})]$ is the equilibrium value of $\Delta ^{\prime }\left(
\alpha ,t\right) $ corresponding to the equilibrium value for a monomer in
the ground state, and we retained only terms that are proportional to $%
|\Omega _{R}(t)|^{2}$ on the right-hand side of Eq.(\ref{eq:rhojjfin_b3}).
The next procedure is similar to that used for obtaining Eq.(\ref{eq:n_j5})
(see above). Integrating Eq.(\ref{eq:rhojjfin_b3}), using Green function,
Eq.(\ref{eq:GF}), we obtain an integral equation that is similar to Eq.(\ref%
{eq:rhojjint2}). Then integrating both sides of the obtained integral
equation with respect to $\alpha ,$ and bearing in mind Eq.(\ref%
{eq:lineshape1}), we get%
\begin{equation}
\frac{dn_{1}}{dt}=-\sigma _{a}(\omega _{21})\tilde{J}(t)\func{Re}\frac{\bar{W%
}_{a}(\omega )}{1-ip\pi W_{a}(\omega )}+\frac{n_{2}}{T_{1}}  \label{eq:n_j7}
\end{equation}%
where the term $\func{Re}\{\bar{W}_{a}(\omega )/[1-ip\pi W_{a}(\omega )]\}$
describes the absorption spectrum of molecules susceptible to the
dipole-dipole intermolecular interactions expressed through their monomer
spectra $W_{a}$. This term agrees with the coherent exciton scattering (CES)
approximation \cite{Briggs02CP,Briggs06CP,Briggs06PRL}. The latter is well
suited to describe absorption spectra of both J- and H-aggregates using
their monomer spectra and the intermolecular interaction strength that is a
fitting parameter.

It is worthy to note that the term $1/[1-ip\pi W_{a}(\omega )]$ on the
right-hand side of Eq.(\ref{eq:n_j7}) amounts to the Pade approximant [0/1] 
\cite{Bak81} that is the sum of diagrams of a certain type \cite%
{Saibatalov_Fainberg87}. Indeed, the term under discussion is the sum of the
infinite geometrical series%
\begin{equation}
\frac{1}{1-ip\pi W_{a}(\omega )}=\dsum\limits_{m=0}^{\infty }p^{m}[i\pi
W_{a}(\omega )]^{m}  \label{eq:Born_series}
\end{equation}%
where the right-hand side of Eq.(\ref{eq:Born_series}) multiplied by $%
W_{a}(\omega )$ may be considered as a Born series with the interaction
parameter $p$.

\subsubsection{Description of the absorption of J-aggregates}

Applying expression $\func{Re}\{\bar{W}_{a}(\omega )/[1-ip\pi W_{a}(\omega
)]\}$ to the description of the absorption of J-aggregates, one should take
into account that the Gaussian shape of the monomer absorption spectrum
obtained in the \textquotedblright slow modulation\textquotedblright\ limit
is correct only near the absorption maximum. The wings decline much slower
as $\left( \omega _{21}-\omega \right) ^{-4}$ \cite{Rautian67}. At the same
time, the expression under discussion has a pole, giving strong absorption,
when $1/(p\pi )=-\func{Im}W_{a}(\omega ).$ If parameter of the dipole-dipole
intermolecular interaction $p$ is rather large, the pole may be at a large
distance from the absorption band maximum where the \textquotedblright slow
modulation\textquotedblright\ limit breaks down. This means one should use
exact expression for the monomer spectrum $W_{a}$ that is not limited by the
\textquotedblright slow modulation\textquotedblright\ approximation, and
properly describe both the central spectrum region and its wings. The exact
calculation of the vibrationally equilibrium monomer spectrum for the
Gaussian-Markovian modulation with the exponential correlation function $%
S(t)=\exp (-|t|/\tau _{s})$ gives \cite{Rautian67,Fai85} (see Appendix)

\begin{equation}
W_{a}(\omega )=\frac{\tau _{s}}{\pi }\frac{\Phi (1,1+x_{a};\sigma _{2s}\tau
_{s}^{2})}{x_{a}}  \label{eq:W_af,exp}
\end{equation}%
where $x_{a}=\tau _{s}/(2T_{1})+\sigma _{2s}\tau _{s}^{2}+i\tau _{s}(\omega
_{21}-\omega )$, $\Phi (1,1+x_{a};\sigma _{2s}\tau _{s}^{2})$ is a confluent
hypergeometric function \cite{Abr64}.

\FRAME{ftbpFU}{4.1277in}{2.0366in}{0pt}{\Qcb{Absorption spectra (in terms of 
$\protect\tau _{s}/\protect\pi $) of the J-aggregate (solid line) and the
corresponding monomer (dashed line) in the case of slow modulation ($\protect%
\sqrt{\protect\sigma _{2s}}\protect\tau _{s}=10.9>>1$). Dimensionless
parameter is $\Delta =\protect\tau _{s}(\protect\omega _{21}-\protect\omega %
) $.}}{\Qlb{fig:abs1}}{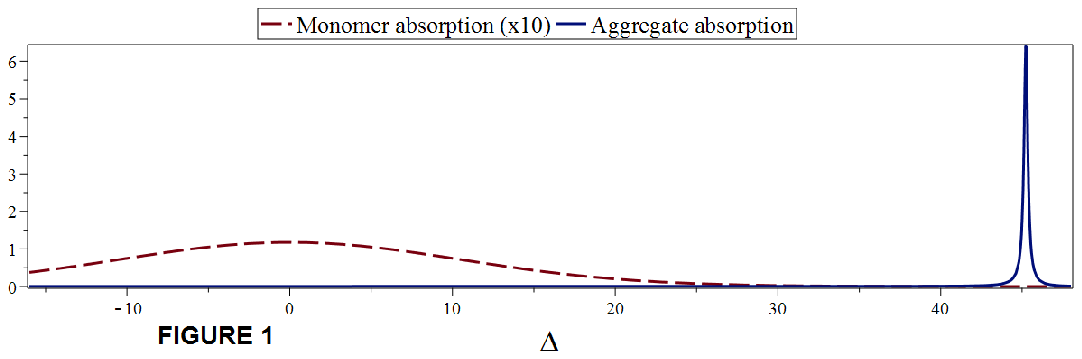}{\special{language "Scientific Word";type
"GRAPHIC";display "USEDEF";valid_file "F";width 4.1277in;height
2.0366in;depth 0pt;original-width 8.1976in;original-height 3.3477in;cropleft
"0";croptop "1";cropright "1";cropbottom "0";filename '../Submitting
1/abs.eps';file-properties "XNPEU";}} 
Figs. \ref{fig:abs1}, \ref{fig:abs1a} and \ref{fig:abs2} show the
calculation results of the absorption spectra of J-aggregates according to
the expression $\func{Re}\{W_{a}(\omega )/[1-ip\pi W_{a}(\omega )]\}$ on the
right-hand side of Eq.(\ref{eq:n_j7}) and Eq.(\ref{eq:W_af,exp})$,$ and
their comparison with the monomer spectra $\func{Re}W_{a}(\omega )$. The
spectra of Fig.\ref{fig:abs1} correspond to a dense collection of molecules (%
$N=10^{21}cm^{-3}$, \ Ref.\cite{Noginov13APL}) with parameters close to
those of molecule LD690 \cite{Fai98}: $\sqrt{\sigma _{2s}}=546$ $cm^{-1}$, $%
\tau _{s}=10^{-13}s$, $D_{12}=10^{-17}$ CGSE that gives $\omega _{st}=1420$ $%
cm^{-1}$,\ $p=\allowbreak 2107$ $cm^{-1}$. We put $T_{1}=10^{-9}s$.

One can see that in spite of strong narrowing the J-aggregate spectra with
respect to those of monomers, the vibrations still give rather important
contribution to broadening the J-aggregate spectra that may be crucial.
Indeed, the half bandwidth of the J-aggregate absorption spectrum is about $%
3\cdot 10^{12}$ $rad/s$ that may far exceed the lifetime contribution. So,
disregarding vibrations in the description of the J-aggregate spectra may be
incorrect. Moreover, above parameters for molecule LD690 in methanol were
obtained using only LFOA vibrations $\{\omega _{s}\}$ for the simulation of
its spectra. If one in addition uses also high frequency OA intramolecular
vibrations (like C-C $\sim $ 1400 $cm^{-1}$) for the simulation (see below),
then the second central moment $\sigma _{2s}$ should be related rather to a
vibronic transition with respect to the high frequency OA vibration than to
the whole spectrum, i.e. the value of $\sigma _{2s}$ diminishes. Fig.\ref%
{fig:abs1a} shows absorption spectra of the J-aggregate and the
corresponding monomer when parameter $\sqrt{\sigma _{2s}}\tau _{s}=3.16$ is
smaller than that for Fig.\ref{fig:abs1}. One can see lesser narrowing the
J-aggregate spectrum with respect to that of a monomer. In contrast, the
J-aggregate absorption spectrum calculated using the monomer spectrum $W_{a}$%
, Eq.(\ref{eq:W}), and, as a consequence, the Gaussian absorption spectrum,
Eq.(\ref{eq:abs}), is extremely narrow (see also \cite{Briggs06PRL}).\FRAME{%
ftbpFU}{3.1528in}{3.3168in}{0pt}{\Qcb{Absorption spectra (in terms of $%
\protect\tau _{s}/\protect\pi $) of the J-aggregate (solid line) and the
corresponding monomer (dashed line) for $\protect\sqrt{\protect\sigma _{2s}}%
\protect\tau _{s}=3.16$ and $p\protect\tau _{s}=5$.}}{\Qlb{fig:abs1a}}{%
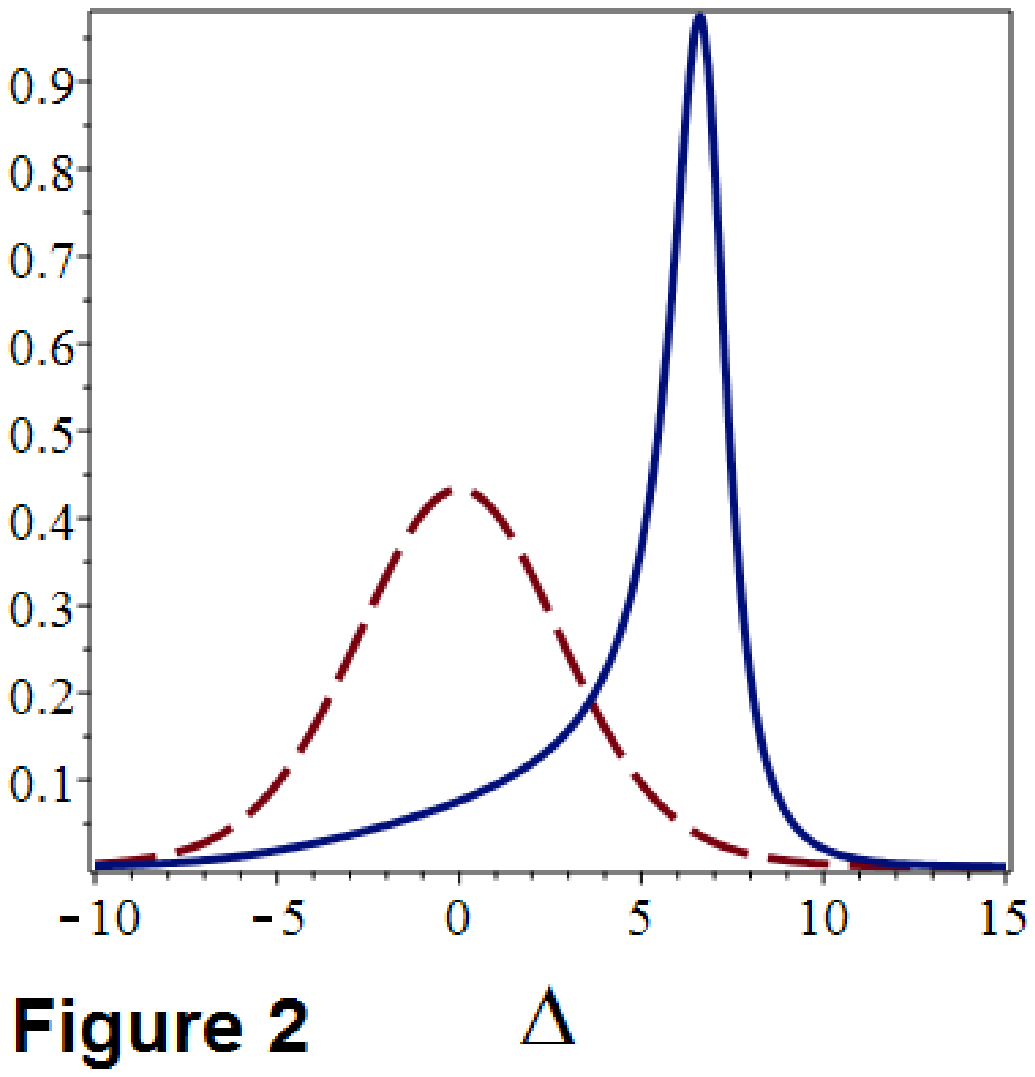}{\special{language "Scientific Word";type
"GRAPHIC";maintain-aspect-ratio TRUE;display "USEDEF";valid_file "F";width
3.1528in;height 3.3168in;depth 0pt;original-width 7.4988in;original-height
7.8923in;cropleft "0";croptop "1";cropright "1";cropbottom "0";filename
'../Submitting 1/abs1a.eps';file-properties "XNPEU";}}

\FRAME{ftbpFU}{2.8712in}{3.0557in}{0pt}{\Qcb{Absorption spectra (in terms of 
$\protect\tau _{s}/\protect\pi $) of the J-aggregate (solid line) and the
corresponding monomer (dashed line) in the case of fast modulation ($\protect%
\sqrt{\protect\sigma _{2s}}\protect\tau _{s}=0.1<<1$) for $p\protect\tau %
_{s}=0.3$.}}{\Qlb{fig:abs2}}{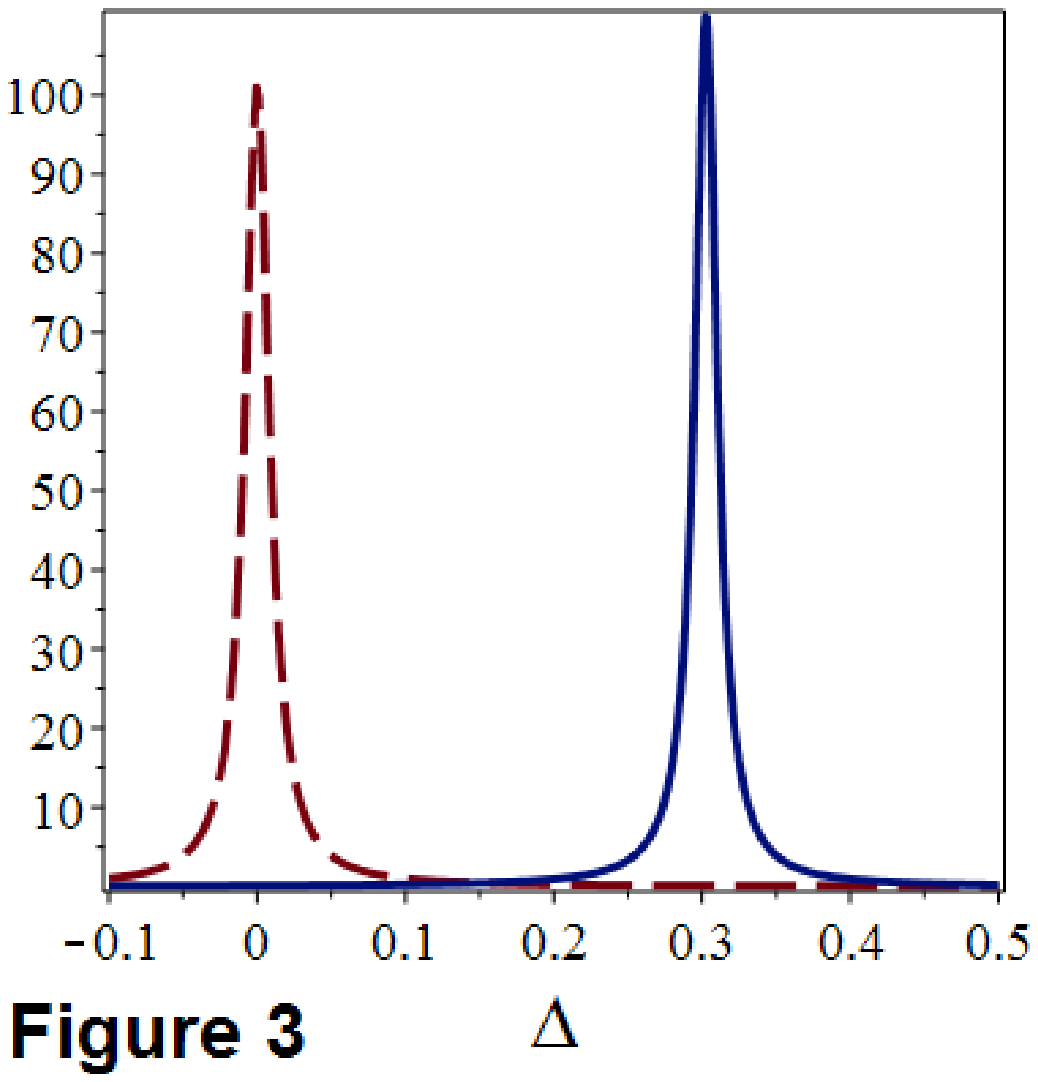}{\special{language "Scientific
Word";type "GRAPHIC";maintain-aspect-ratio TRUE;display "USEDEF";valid_file
"F";width 2.8712in;height 3.0557in;depth 0pt;original-width
7.1079in;original-height 7.5688in;cropleft "0";croptop "1";cropright
"1";cropbottom "0";filename '../Submitting 1/abs2.eps';file-properties
"XNPEU";}}

For fast modulation when $\sigma _{2s}\tau _{s}^{2}<<1$, the aggregate
spectrum only shifts with respect to the monomer one almost without changing
its shape (Fig.\ref{fig:abs2}). Indeed, $\Phi (1,1+x_{a(f)};\sigma _{2s}\tau
_{s}^{2})\approx 1$\ for $\sigma _{2s}\tau _{s}^{2}<<1$. In that case $%
W_{a}(\omega )\approx (\tau _{s}/\pi )/x_{a}$, and the term $\func{Re}\{\bar{%
W}_{a}(\omega )/[1-ip\pi W_{a}(\omega )]\}$ on the right-hand side of Eq.(%
\ref{eq:n_j7}) becomes

\begin{eqnarray}
\func{Re}\frac{W_{a}(\omega )}{1-ip\pi W_{a}(\omega )} &\approx &\frac{1}{%
\pi }\func{Re}\frac{1}{\frac{1}{2T_{1}}+\sigma _{2s}\tau _{s}+i(\omega
_{21}-\omega -p)}  \notag \\
&=&W_{a}(\omega +p)  \label{eq:abs_hom_broada}
\end{eqnarray}%
In other words, if the monomer spectrum has Lorentzian shape, the aggregate
spectrum is simply shifted monomer spectrum. In that case both the approach
based on the self-energy depending on the effective vibrational coordinate,
and the approach based on the population difference ("lesser" GFs) depending
on the effective vibrational coordinate give the same absorption spectrum of
molecules susceptible to the dipole-dipole intermolecular interactions.

\section{Description of the absorption of H-aggregates}

\label{sec:H}

Applying expression $\func{Re}\{W_{a}(\omega )/[1-ip\pi W_{a}(\omega )]\}$
(see Section\ref{sec:lineshape}) to the description of the absorption of
H-aggregates, one should take into account also high frequency OA
intramolecular vibrations \cite{Briggs06CP}, in addition to the LFOA
vibrations $\{\omega _{s}\}$ under consideration in our paper. The general
form of Eq.(\ref{eq:n_j7}) enables us to do this. We will consider one
normal high frequency intramolecular oscillator of frequency $\omega _{0}$
whose equilibrium position is shifted under electronic transition. Its
characteristic function $f_{\alpha M}(t)$ is determined by the following
expression \cite{Lin68,Fai00JCP}:

\begin{eqnarray}
f_{\alpha M}(t) &=&\exp (-S_{0}\coth \theta _{0})\sum_{k=-\infty }^{\infty
}I_{k}(S_{0}/\sinh \theta _{0})  \notag \\
&&\times \exp [k(\theta _{0}+i\omega _{0}t)]  \label{eq:charfunMosc}
\end{eqnarray}%
where $S_{0}$ is the dimensionless parameter of the shift, $\theta
_{0}=\hbar \omega _{0}/(2k_{B}T)$, $I_{n}(x)$ is the modified Bessel
function of first kind \cite{Abr64}. Then $W_{a}(\omega )$ can be written as 
\begin{eqnarray*}
W_{a}(\omega ) &=&\frac{1}{\pi }\int_{0}^{\infty }f_{\alpha M}^{\ast
}(t)\exp [i(\omega -\omega _{21})t+g_{s}(t)]dt \\
&=&\sum_{k=-\infty }^{\infty }\frac{\exp (-S_{0}\coth \theta _{0}+k\theta
_{0})}{\pi }I_{k}(\frac{S_{0}}{\sinh \theta _{0}}) \\
&&\times \int_{0}^{\infty }\exp [i(\omega -k\omega _{0}-\omega
_{21})t+g_{s}(t)]dt
\end{eqnarray*}%
where $g_{s}(t)$ is given by Eq.(\ref{eq:g(t)GM}) of the Appendix.
Integrating with respect to $t$, one gets

\begin{eqnarray}
W_{a}(\omega ) &=&\frac{\tau _{s}}{\pi }\exp (-S_{0}\coth \theta
_{0})\sum_{k=-\infty }^{\infty }I_{k}(\frac{S_{0}}{\sinh \theta _{0}}) 
\notag \\
&&\times \exp (k\theta _{0})\frac{\Phi (1,1+x_{ak};\sigma _{2s}\tau _{s}^{2})%
}{x_{ak}}  \label{eq:W_af,expHF}
\end{eqnarray}%
where $x_{ak}=\tau _{s}/(2T_{1})+\sigma _{2s}\tau _{s}^{2}+i\tau _{s}(\omega
_{21}-\omega +k\omega _{0})$. Eq.(\ref{eq:W_af,expHF}) is the extension of
Eq.(\ref{eq:W_af,exp}) to the presence of high frequency intramolecular
vibrations. For $\theta _{0}>>1$ we obtain%
\begin{equation}
W_{a}(\omega )=\frac{\tau _{s}}{\pi }\exp (-S_{0})\sum_{k=0}^{\infty }\frac{%
S_{0}^{k}}{k!}\frac{\Phi (1,1+x_{ak};\sigma _{2s}\tau _{s}^{2})}{x_{ak}}
\label{eq:W_af,expHFLowTemp}
\end{equation}%
If the behavior of the wings of spectra is unimportant, one can use the
\textquotedblright slow modulation\textquotedblright\ limit. In that case
Eq.(\ref{eq:W_af,expHFLowTemp}) may be replaced by 
\begin{equation}
W_{a}(\omega )=\frac{\exp (-S_{0})}{\sqrt{2\pi \sigma _{2s}}}%
\sum_{k=0}^{\infty }\frac{S_{0}^{k}}{k!}w(\frac{\omega -\omega _{21}-k\omega
_{0}}{\sqrt{2\sigma _{2s}}})  \label{eq:W_af,expHFslow}
\end{equation}%
\ \ 

\subsection{Explanation of introducing the empiric red shift of the
absorption spectra of H-aggregates calculated in CES approximation}

It is worthy to note that the CES approximation describes well the shape of
the absorption spectra of H-aggregates. However, the spectra calculated in
the CES approximation are blue shifted with respect to experimental ones 
\cite{Briggs06CP}. To resolve the problem, the authors of Ref.\cite%
{Briggs06CP} empirically introduced additional red shift that could be
understood in the context of our more general theory as the red shift due to
the CMLL mechanism. Indeed, let us write down Eq.(\ref{eq:rho21}) when both
the self-energy ($\sim \tilde{\rho}_{21}$) and the population difference
depends on the effective vibrational coordinate

\begin{eqnarray}
&&\frac{\partial }{\partial t}\tilde{\rho}_{21}\left( \alpha ,t\right) 
\notag \\
&\approx &-i\left( \omega _{21}-\omega -p_{1}-\alpha \right) \tilde{\rho}%
_{21}\left( \alpha ,t\right)  \notag \\
&\approx &i[\frac{\mathbf{D}_{21}\cdot \mathbf{E}\left( t\right) }{2\hbar }%
+p_{2}\dint \tilde{\rho}_{21}\left( \alpha ,t\right) d\alpha ]\rho
_{11}^{\left( 0\right) }\left( \alpha \right)  \label{eq:rho21two_mechanisms}
\end{eqnarray}%
Using the procedure described in Ref.\cite{Fai98}, we get an equation
similar to Eq. (\ref{eq:rhojjfin_b3}) (together with Eq.(\ref%
{eq:rho21integrated_eq})) with the only difference that $\omega _{21}$
should be replaced by $\omega _{21}-p_{1}$, and $p$ - by $p_{2}$%
\begin{eqnarray}
\frac{\partial \rho _{11}\left( \alpha ,t\right) }{\partial t} &=&L_{11}\rho
_{11}\left( \alpha ,t\right) -  \notag \\
&&-\frac{\frac{\pi }{2}\rho _{11}^{\left( 0\right) }\left( \alpha \right)
|\Omega _{R}(t)|^{2}\delta \left( \omega _{21}-\omega -p_{1}-\alpha \right) 
}{|1+p_{2}\int d\alpha \rho _{11}^{\left( 0\right) }\left( \alpha \right)
\zeta (\omega +p_{1}-\omega _{21}+\alpha )|^{2}}  \label{eq:rho11equilibrium}
\end{eqnarray}%
Then similar to Eq.(\ref{eq:n_j7}), we obtain%
\begin{equation}
\frac{dn_{1}}{dt}=-\sigma _{a}(\omega _{21})\tilde{J}(t)\func{Re}\frac{\bar{W%
}_{a}(\omega +p_{1})}{1-ip_{2}\pi W_{a}(\omega +p_{1})}+\frac{n_{2}}{T_{1}}
\label{eq:n_1both}
\end{equation}%
where the term $\func{Re}\{\bar{W}_{a}(\omega +p_{1})/[1-ip_{2}\pi
W_{a}(\omega +p_{1})]\}$ describes the absorption spectrum of molecules
susceptible to the dipole-dipole intermolecular interactions expressed
through their monomer spectra $W_{a}(\omega +p_{1}),$ Eqs.(\ref%
{eq:W_af,expHF}), (\ref{eq:W_af,expHFLowTemp}) and (\ref{eq:W_af,expHFslow}%
). \FRAME{ftbpFU}{2.4146in}{2.4924in}{0pt}{\Qcb{Absorption spectra (in terms
of $\protect\tau _{s}/\protect\pi $) of the H-aggregate (solid line), the
corresponding monomer (dash line) and the H-aggregate without the
contribution of the CMLL mechanism (dash dot line) for $p_{1}=500$ $cm^{-1}$
and $p_{2}=-\allowbreak 1500$ $cm^{-1}$ Dimensionless parameter is $\Delta =%
\protect\tau _{s}(\protect\omega _{21}-\protect\omega )$.}}{\Qlb{fig:absH}}{%
absh.eps}{\special{language "Scientific Word";type
"GRAPHIC";maintain-aspect-ratio TRUE;display "USEDEF";valid_file "F";width
2.4146in;height 2.4924in;depth 0pt;original-width 7.3542in;original-height
7.5922in;cropleft "0";croptop "1";cropright "1";cropbottom "0";filename
'../Submitting 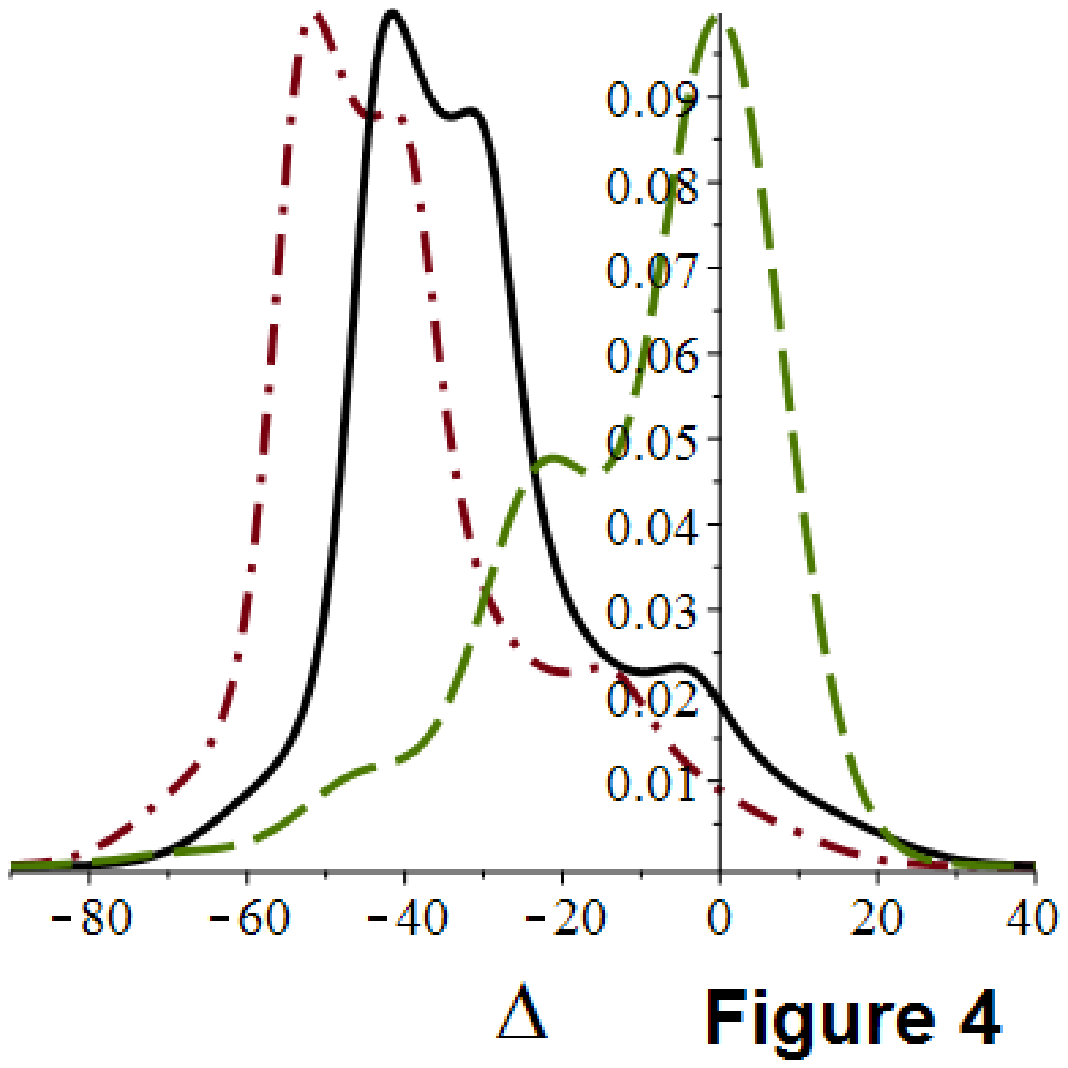';file-properties "XNPEU";}}

Fig.\ref{fig:absH} shows the calculation results of the absorption spectrum
of an H-aggregate according to the expression $\func{Re}\{W_{a}(\omega
+p_{1})/[1-ip_{2}\pi W_{a}(\omega +p_{1})]\}$ on the right-hand side of Eq.(%
\ref{eq:n_1both}) and Eq.(\ref{eq:W_af,expHFLowTemp}) (solid line)$,$ and
its comparison with the monomer spectrum $\func{Re}W_{a}(\omega )$ (dash
line) and the spectrum of H-aggregate, $\func{Re}\{W_{a}(\omega
)/[1-ip_{2}\pi W_{a}(\omega )]\}$, calculated without the contribution of
the CMLL mechanism (dash dot line). The values of parameters are found by
fitting the experimental spectrum of the linear absorption of LD690 in
methanol \cite{Fai02JCP}: $\tau _{s}=10^{-13}s$, $k_{B}T=210$ $cm^{-1}$, $%
\hbar \omega _{st}/(2k_{B}T)=1.99$, $S_{0}=0.454,$ $\omega _{0}=1130$ $%
cm^{-1}$, $\sigma _{2s}=\omega _{st}k_{B}T/\hbar $. The spectra presented in
Fig.\ref{fig:absH} manifest that though the shape of the H-aggregate
spectrum is fully described by the self-energy not depending on the
effective vibrational coordinate, its position (including the additional red
shift of the experimental spectra of H-aggregates \cite{Briggs06CP}) may be
correctly described only taking the CMLL mechanism into account. In other
words, our more general theory enables us to describe both the shape and the
position of the experimental spectra of H-aggregates due to the self-energy
and the population difference ("lesser" GFs) both depending on the effective
vibrational coordinate that leads to their frequency dependence. This can be
understood as follows. The frequency dependent "lesser" GFs corresponding to
the CES approximation describe well the spectral shapes of H-aggregates. The
latters can interact with each other by the dipole-dipole interaction
leading to the CMLL red shift that is described by the frequency dependent
self-energy.

\section{Bistability}

\label{sec:bistability}

We saw in Section\ref{sec:H} that the dependence of both the self-energy and
the population difference on vibrational coordinates manifests in
experiment. Let us concentrate first on the self-energy depending on the
effective vibrational coordinate. The corresponding Eqs. (\ref{eq:rhojjfin_b}%
) and (\ref{eq:n_j5}) for populations are nonlinear equations and can
demonstrate a bistable behavior. We shall consider the CMLL mechanism of the
dipole-dipole intermolecular interactions leading to the red shift (the
first term on the right-hand side of Eq.(\ref{eq:n_j5})) when one can use
Eqs.(\ref{eq:W})-(\ref{eq:abs}) for the monomer spectra. Fig.\ref%
{fig:n2_bistability} shows steady-state solutions of Eq.(\ref{eq:n_j5}) for $%
n_{2}$ as a function of the power density of the exciting radiation $\tilde{J%
}$ at different detunings $\omega _{21}-\omega $. \FRAME{ftbpFU}{2.7616in}{%
4.3578in}{0pt}{\Qcb{(Color online) Dependence of excited state population $%
n_{2}$ on power density of the exciting radiation $\tilde{J}$ at different
detunings $\protect\omega _{21}-\protect\omega $. Dimensionless parameters
are $J=\protect\sigma _{a}\tilde{J}T_{1}$ and $y=(\protect\omega _{21}-%
\protect\omega )/\protect\sqrt{2\protect\sigma _{2s}}$. Parameters $\protect%
\sqrt{\protect\sigma _{2s}}$, $\protect\omega _{st}$ and $p$ are identical
to those of Section\protect\ref{sec:lineshape}. Inset: Equilibrium spectra
of the absorption (Abs) and the emission (Em); the arrows limit the
frequency interval where calculated excited state populations $n_{2}$ show
bistability. }}{\Qlb{fig:n2_bistability}}{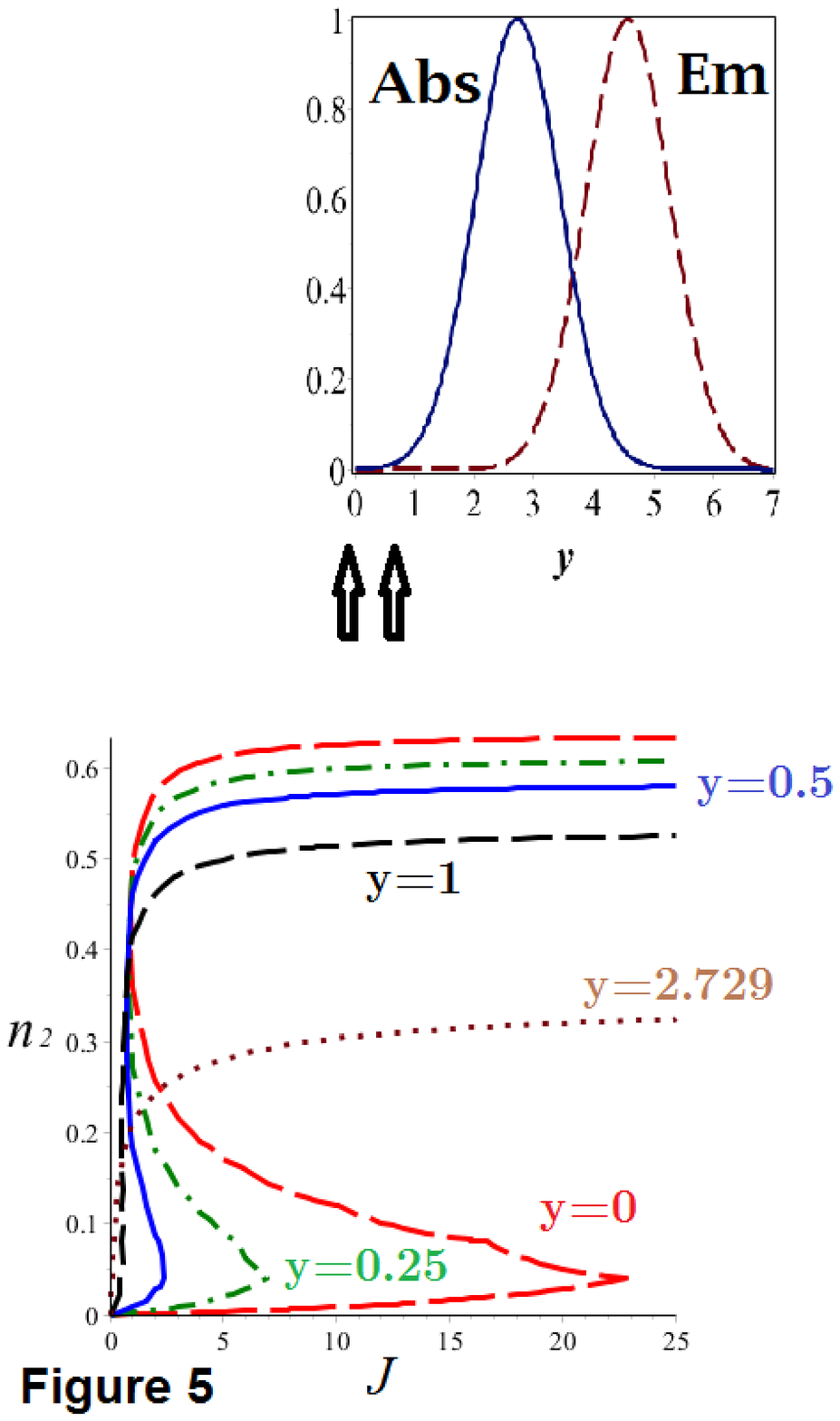}{\special%
{language "Scientific Word";type "GRAPHIC";maintain-aspect-ratio
TRUE;display "USEDEF";valid_file "F";width 2.7616in;height 4.3578in;depth
0pt;original-width 7.0076in;original-height 11.1025in;cropleft "0";croptop
"1";cropright "1";cropbottom "0";filename '../Submitting
1/bistability5a.eps';file-properties "XNPEU";}} 
One can see that each value of $\tilde{J}$ within the corresponding interval
produces three different solutions of Eq.(\ref{eq:n_j5}) for dimensionless
detunings $y=(\omega _{21}-\omega )/\sqrt{2\sigma _{2s}}=0,$ $0.25$ and $0.5$%
, however, only lower and upper branches are stable \cite%
{Bogoliubov-Mitropolski61}. Such detunings correspond to the excitation at
the short-wave part of the equilibrium absorption spectrum (see the Inset to
Fig.\ref{fig:n2_bistability}). As the excited state population increases,
the spectrum exhibits the blue shift (see Eq.(\ref{eq:n_j5})) that should
essentially contribute to the absorption. As a matter of fact, the bistable
behavior of the population arises from the dependence of the resonance
frequency of the molecules in dense medium on the number of excited
molecules. In contrast, larger $y=1$, $2.729$ correspond to the excitation
closer to the central part of the equilibrium absorption spectrum. In that
case the blue shift produces lesser increasing the absorption and even can
decrease it (for $y=2.729$), so that the bistable behavior disappears.

\subsection{Diminishing light intensity necessary for bistablity below the
damage threshold}

Since experiments on bistablity should be done with CW pumping or pulse
pumping when the distance between pulses is smaller than the excited state
lifetime $T_{1}$, the light intensity ought to be smaller than that in usual
pulse experiments with thin films of dyes \cite{Noginov15Photonics} (as to
CW pumping, the damage threshold should be much smaller than $1$ $MW/cm^{2}$ 
\cite{Noginov}). The resolution of the problem may be achieved using
3-electronic states system with the long-living triplet state $T_{1}$ near
(but below) the excited singlet state $S_{1}$, and fast intersystem crossing 
$S_{1}\rightarrow T_{1}$, Fig.\ref{fig:3level}.\FRAME{ftbpFU}{4.058in}{%
2.4923in}{0pt}{\Qcb{3-electronic states system. 1,2 - sinlet states, 3 -
triplet state.}}{\Qlb{fig:3level}}{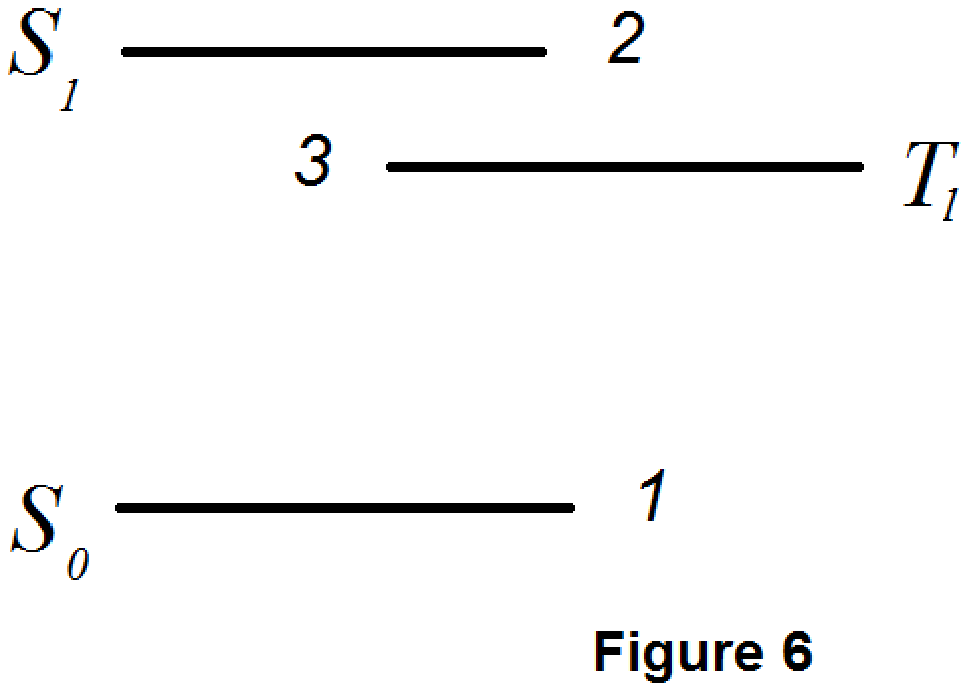}{\special{language "Scientific
Word";type "GRAPHIC";maintain-aspect-ratio TRUE;display "USEDEF";valid_file
"F";width 4.058in;height 2.4923in;depth 0pt;original-width
7.1705in;original-height 4.385in;cropleft "0";croptop "1";cropright
"1";cropbottom "0";filename '../Submitting 1/3level.eps';file-properties
"XNPEU";}} The rose bengal \cite{Cheng-Chung08,Neckers89,Larkin99,Fini07}
and platinum-octaethyl-porphyrin (PtOEP) \cite{Bansal06,Nifiatis11} are
examples. They possess also large absorption cross-section ($\sigma
_{a}\approx 10^{-15}$ $cm^{2}$ for PtOEP) that is necessary for significant
light-induced change of the dielectric function. In such a case one can
achieve strong depletion of the ground state $S_{0}$ using intensities much
smaller than $1$ $MW/cm^{2}$. Indeed, generalizing Eq.(\ref{eq:n_j5}) to the
three-level electronic system and bearing in mind Eqs.(\ref{eq:W}) and (\ref%
{eq:abs}), one gets

\begin{eqnarray}
\frac{dn_{1}}{dt} &=&\sigma _{a}\exp [-\frac{(\omega _{21}-p\Delta n-\omega
-\omega _{st})}{2\sigma _{2s}}^{2}]\tilde{J}\{n_{2}-  \notag \\
&&-n_{1}\exp \left[ \hbar \beta \left( \omega +p\Delta n-\omega _{21}+\frac{%
\omega _{st}}{2}\right) \right] \}  \notag \\
&&+\frac{n_{2}}{T_{1}}+w_{31}n_{3}  \label{eq:dn1/dt}
\end{eqnarray}%
\begin{eqnarray}
\frac{dn_{2}}{dt} &=&-\sigma _{a}\exp [-\frac{(\omega _{21}-p\Delta n-\omega
-\omega _{st})}{2\sigma _{2s}}^{2}]\tilde{J}\{n_{2}-  \notag \\
&&-n_{1}\exp \left[ \hbar \beta \left( \omega +p\Delta n-\omega _{21}+\frac{%
\omega _{st}}{2}\right) \right] \}  \notag \\
&&-n_{2}(\frac{1}{T_{1}}+w_{23})  \label{eq:dn2/dt}
\end{eqnarray}%
\begin{equation}
\frac{dn_{3}}{dt}=n_{2}w_{23}-w_{31}n_{3}  \label{eq:dn3/dt}
\end{equation}%
where $w_{ij}$ is the probability of radiationless transition $i\rightarrow
j $, and $n_{1}+n_{2}+n_{3}=1$.\FRAME{ftbpFU}{2.3926in}{2.4923in}{0pt}{\Qcb{%
Dependence of the ground state population $n_{1}$ on power density of the
exciting radiation $\tilde{J}$ at dimensionless detuning $y=(\protect\omega %
_{21}-\protect\omega )/\protect\sqrt{2\protect\sigma _{2s}}=0.5$ for $%
T_{1}=10^{-9}$ $s$. Probabilities of radiationless transitions, $%
w_{23}=10^{10}$ $s^{-1}$ and $w_{31}=10^{4}$ $s^{-1}$, are close to those in
rose bengal \protect\cite{Cheng-Chung08,Neckers89,Larkin99,Fini07} and
platinum-octaethyl-porphyrin (PtOEP) \protect\cite{Bansal06,Nifiatis11}.
Parameters $\protect\sqrt{\protect\sigma _{2s}}$, $\protect\omega _{st}$ and 
$p$ are identical to those of Section\protect\ref{sec:lineshape}. }}{\Qlb{%
fig:n1_bistability3levels}}{bist3level.eps}{\special{language "Scientific
Word";type "GRAPHIC";maintain-aspect-ratio TRUE;display "USEDEF";valid_file
"F";width 2.3926in;height 2.4923in;depth 0pt;original-width
6.9819in;original-height 7.2751in;cropleft "0";croptop "1";cropright
"1";cropbottom "0";filename '../Submitting 1/bist3level.eps';file-properties
"XNPEU";}}

Fig.\ref{fig:n1_bistability3levels} shows the steady-state solution of Eqs.(%
\ref{eq:dn1/dt}), (\ref{eq:dn2/dt}) and (\ref{eq:dn3/dt}) for $n_{1}$ as a
function of the power density of the exciting radiation $\tilde{J}$ for $%
y=(\omega _{21}-\omega )/\sqrt{2\sigma _{2s}}=0.5$. One can see that the
bistable behavior persists for the 3-electronic states system, however,
involves much smaller light intensities than those for 2-electronic states
system without triplet $T_{1}$ (see Fig.\ref{fig:n2_bistability}). In
addition, rose bengal is characterized by gigantic third order
susceptibility $\chi ^{(3)}\sim 10^{-3}$ $esu$.

\section{Conclusion}

\label{sec:conclusion}

In this work we have developed a mean-field theory of light-induced optical
properties of photonic organic materials taking the collective effects into
account. Our consideration is based on the model of the interaction of
strong shaped laser pulse with organic molecules, Refs.\cite%
{Fai98,Fai00JCP,Fai02JCP}, extended to the dipole-dipole intermolecular
interactions in the condensed matter. We show that such a generalization can
describe both a red shift of the resonance frequency of isolated molecules,
according to the CMLL mechanism \cite{Klein-Furtak86}, and the wide
variations of their spectra related to the aggregation of molecules into J-
or H-aggregates. In particular case of weak radiation we recover the CES
approximation \cite{Briggs02CP,Briggs06CP,Briggs06PRL}. We show that the
experimental absorption spectra of H-aggregates may be correctly described
only if one takes both mechanisms into account. Our theory contains
experimentally measured quantities that makes it closely related to
experiment.

The bistable response of organic materials in the condensed phase has been
demonstrated using the electron-vibrational model. We have shown that using
molecules with long-living triplet state $T_{1}$ near excited singlet state $%
S_{1}$, and fast intersystem crossing $S_{1}\rightarrow T_{1}$ enables us to
diminish CW light intensity needed for observing bistability below the
damage threshold of thin organic films.

The phenomenon of bistability in spatially distributed systems can result in
the generation of the switching waves in photonic organic materials \cite%
{Fainberg17APL}.

I thank M. A. Noginov for useful discussions and D. Huppert who attracted my
attention to molecules Rose Bengal and platinum-octaethyl-porphyrin (PtOEP).

\section{Appendix}

In the case of the Gaussian modulation of the electronic transition by the
vibrations, the absorption lineshape is given by \cite%
{Kub62R,Fai85,Muk95,Fai03AMPS}%
\begin{equation}
F_{a}(\omega )=\frac{1}{\pi }\func{Re}\int_{0}^{\infty }\exp [i(\omega
-\omega _{21})t+g(t)]dt  \label{eq:fl}
\end{equation}%
where 
\begin{equation}
g(t)=-\int_{0}^{t}dt^{\prime }(t-t^{\prime })K(t^{\prime })  \label{eq:g(t)}
\end{equation}%
is the logarithm of the characteristic function of the spectrum of
single-photon absorption after substraction of a term which is linear with
respect to $t$ and determines the first moment of the spectrum, $K(t)$ is
the correlation function. Then%
\begin{equation}
W_{a}(\omega )=\frac{1}{\pi }\int_{0}^{\infty }\exp [i(\omega -\omega
_{21})t+g(t)]dt  \label{eq:W_af}
\end{equation}

For the exponential correlation function $K_{s}(t)=\sigma _{2s}\exp
(-|t|/\tau _{s})$, we get%
\begin{equation}
g_{s}(t)=-\sigma _{2s}\tau _{s}^{2}[\exp (-t/\tau _{s})+\frac{t}{\tau _{s}}%
-1]  \label{eq:g(t)GM}
\end{equation}%
that leads to Eq.(\ref{eq:W_af,exp}) of section \ref{sec:lineshape}.

\providecommand{\refin}[1]{\\ \textbf{Referenced in:} #1}

\end{document}